\documentclass{pj}

\usepackage{graphicx,epsfig,subfigure}

\def\e{\begin{equation}}
\def\f{\end{equation}}

\begin{document}\setcounter{page}{1}

\pjheader{Vol.\ x, y--z, 2004}

\title{Grounded Uniaxial Material Slabs as Magnetic Conductors}


\author{O.~Luukkonen, C.~R.~Simovski, and S.~A.~Tretyakov}
\address{Department of Radio Science and Engineering/SMARAD CoE \\TKK Helsinki
University of Technology \\P.O. 3000, FI-02015 TKK, Finland\\(email:olli.luukkonen@tkk.fi,csimovsk@cc.hut.fi,sergei.tretyakov@tkk.fi)}
\runningauthor{Luukkonen, Simovski, and Tretyakov}

\tocauthor{O.~Luukkonen: olli.luukkonen@tkk.fi}

\begin{abstract} The objective of this paper is all-angle artificial magnetic conductor, i.e. artificial magnetic conductor that has stable magnetic-wall effect with respect to the incidence angle. Furthermore, we seek for a design that would be easy for manufacturing. In order to achieve this we use grounded uniaxial material slabs and we do not constrict ourselves to naturally available materials. Instead, we assume that the desired parameters can be synthesized using the emerging artificial electromagnetic materials. It is found that it is possible to have an all-angle magnetic-wall effect for both TE and TM polarization. Especially for the TM fields the structure would be easily manufacturable. The proposed structure has similar appearance as more well-known artificial impedance surfaces, but the design parameters and the physical properties behind the magnetic wall effect are novel. The performance of the proposed artificial magnetic conductor is verified with numerical simulations. This paper introduces a new approach how to obtain a magnetic-wall effect. It is possible to use this this approach also together with other ways of obtaining the magnetic-wall effect for dual-band operation.
\end{abstract}




\section{Introduction}\label{Introduction}
Artificial magnetic conductors (AMC) have been studied widely in the literature after the seminal paper of D.~Sievenpiper et al. published in 1999 \cite{Sievenpiper}. Especially, the use of these surfaces for lowering the height of horizontal wire antennas has attracted much attention. In \cite{Sievenpiper, Monorchio,  Clavijo, Kern, feresidis, Simovski1,Goussetis,Luukkonen1, Luukkonen2} artificial magnetic conductors have been realized as a certain resonant structure, not involving resonant magnetically or electrically polarizable particles. In the vicinity of the structure resonance the surface impedance becomes very high in the absolute value, making the use of these surfaces as magnetic conductors possible. In corrugated surfaces the quarter-wavelength resonances can be utilized in a similar way (see e.g. \cite{Kildal}). In addition to resonant structures, also more conventional materials (in the sense of magnetic or electric polarizations excited by magnetic or electric stimuli) have been studied \cite{Ziolkowski, Erentok, Mervi, Mervi2}. In these cases the materials have been considered without paying attention to the possible anisotropy. In \cite{Mervi} it was shown that the radiation properties of a horizontal dipole can be improved by a substrate implemented as a grounded slab with nearly zero permittivity. Such a substrate operates as a magnetic conductor for all angles of incidence of TM-polarized electromagnetic waves. 

The uniaxial symmetry of the structure is the most general allowed symmetry, if the magnetic-wall effect should be isotropic in the surface plane. Thus, we will study the most general approach to the realization of the all-angle magnetic walls using material layers. The fully isotropic layers are a special case covered by this theory, but we expect that allowing uniaxial anisotropy will make the actual realization much simpler. Indeed, to engineer, for example, passive $\varepsilon \approx 0$ materials one needs inclusions which are inherently anisotropic. Naturally, it is possible to manufacture isotropic materials from anisotropic inclusions by grouping and orienting the particles in such a way that an electric or a magnetic stimuli along all of the orthogonal direction will create the same response. This is, however, a troublesome and most likely impractical way, if electrically thin realizations are needed. The possibility to avoid the requirement of full 3D isotropy would greatly simplify the structure and its manufacturing. 

In this paper we systematically study what uniaxial material property can allow us to realize all-angle magnetic conductors. We will not restrict ourselves to the naturally available materials but assume that the desired parameters can be synthesized using emerging metamaterials (e.g., \cite{extreme}). In a uniaxial material, the relative permittivity or permeability have different values along the transversal and normal axes (defined with respect to the surface of the material slab). Uniaxial materials provide hence an extra degree of freedom compared to the conventional materials. As we will see, the use of artificial materials with extreme parameters offers a possibility for design of structures with superior performance, not achievable with conventional materials. 

In the following section II we derive the needed analytical models for the analysis of the grounded uniaxial material slabs. In sections III and IV we use these models to study what kind of material parameter values are needed for all-angle magnetic conductors for the TM and TE fields, respectively. The study reveals that for the TM fields it is sufficient to have just the normal component of the permittivity to be close to zero. This can be done by using wire medium at its plasma frequency and suppressing its spatial dispersion. In section V we will discuss the issue of suppression of the spatial dispersion in the wire medium. An example of the structure is given in section VI in which we also verify our result by simulations.

\section{Surface impedance of a grounded uniaxial material slab}\label{Surface impedance of a grounded uniaxial material slab}

\begin{figure}[t!]
\centering \epsfig{file = 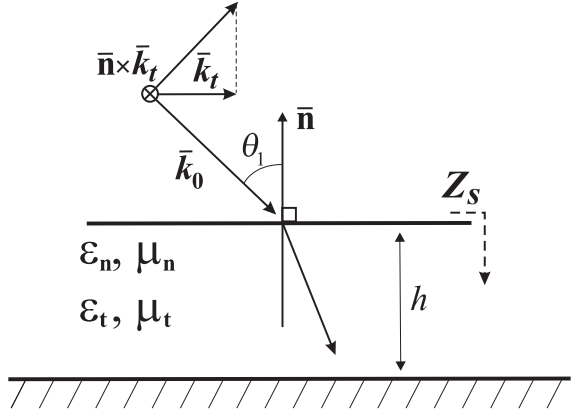, width = 7.5cm}
\caption{ An illustrative picture of the grounded uniaxial material
slab. Index $t$ corresponds to the direction tangential to the interfaces, and index $n$ marks the components
along the unit vector $\mathbf n$.} \label{fig:1}
\end{figure}

\begin{figure}[t!]
\centering \epsfig{file = 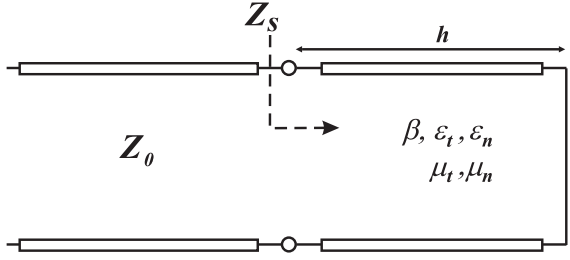, width = 7.5cm}
\caption{ Transmission-line model for the grounded uniaxial material
slab.} \label{fig:2}
\end{figure}

In Fig.~\ref{fig:1} a schematic picture of a grounded uniaxial
material slab is shown. The properties of the material slab are
described by the effective transverse and normal permittivities
($\varepsilon_{\rm t}$ and $\varepsilon_{\rm n}$) and permeabilities
($\mu_{\rm t}$ and $\mu_{\rm n}$), respectively. It is convenient to
model the structure shown in Fig.~\ref{fig:1} with a surface
impedance at the top interface of the structure. This surface
impedance can be calculated through the transmission-line model
illustrated by Fig.~\ref{fig:2} \cite{Luukkonen2}. For this we need both the wave impedance and the normal component $\beta$ of the wave number in the uniaxial material of the substrate.

The wave impedances in the uniaxial material read for TE- and
TM-polarized incident plane waves, respectively, as
\cite{Tretyakov}: \e Z^{\rm TE} = \frac{\omega\mu_{\rm
t}}{\beta_{\rm TE}}\label{eq:Z^TE},\f \e Z^{\rm TM} =
\frac{\beta_{\rm TM}}{\omega\varepsilon_{\rm t}}, \label{eq:Z^TM}\f
Here, the
normal components of the wave vectors in the uniaxial slab read
\cite{Tretyakov}:
\e \beta_{\rm TE}^2 = \omega^2\varepsilon_{\rm
t}\mu_{\rm t} - k_{\rm t}^2\frac{\mu_{\rm t}}{\mu_{\rm
n}}\label{eq:beta_TE},\f \e \beta_{\rm TM}^2 =
\omega^2\varepsilon_{\rm t}\mu_{\rm t} - k_{\rm
t}^2\frac{\varepsilon_{\rm t}}{\varepsilon_{\rm
n}}\label{eq:beta_TM}, \f
where $k_{\rm t}$ is the transverse wave number.
Using the transmission-line model in
Fig.~\ref{fig:2} and \ref{eq:Z^TE}--\ref{eq:beta_TM}, we can
write the surface impedances for both polarizations as (see also
\cite{Ikonen}): \e Z_{\rm s}^{\rm TE} = j\frac{\omega\mu_{\rm
t}}{\beta_{\rm TE}}\tan\left(\beta_{\rm TE}h\right),
\label{eq:Z_s^TE}\f \e Z_{\rm s}^{\rm TM} = j\frac{\beta_{\rm
TM}}{\omega\varepsilon_{\rm t}}\tan\left(\beta_{\rm TM}h\right).
\label{eq:Z_s^TM}\f In order to realize a magnetic conductor, the
absolute values of these expressions need to be maximized.

Let us next look more closely into the cases of TM and TE
polarizations separately. Our goal for both polarizations is to
realize a magnetic conductor whose performance would not be affected
by the incident angle nor by the finite electrical thickness of the
slab.

\section{Magnetic conductors for TM polarization}\label{Magnetic conductors for TM polarization}

The main problems for artificial magnetic conductors are the
frequency and angular dependencies of the magnetic wall effect. A
grounded dielectric slab with the quarter-wavelength electrical
thickness serves as a good example here. The magnetic wall effect
due to the $\lambda/4$-thickness of the slab is clearly highly
dependent on the frequency. In the case of low relative permittivity
of the substrate the effect is also highly dependent on the
incidence angle. The magnetic wall effect moves to higher
frequencies as the incidence angle grows. It is desirable to decrease both frequency and angular dependencies of the magnetic wall effect.

We first study the properties
of uniaxial material slabs for which the transverse and normal
components of the relative permeability are both positive
and the components of the permittivity are arbitrary chosen. We
will then study the inverse case when the transverse component of the permeability is arbitrary chosen. For simplicity of the analysis we neglect the
losses in the structure.

Let us first consider reducing the periodical dependency of the
surface impedance on the frequency by designing an imaginary
$\beta_{\rm TM}$. This would mean that the waves propagating along
the normal to the surface in the uniaxial material slab would be
evanescent. In the case when $\varepsilon_{\rm n} \rightarrow (+0)$
we approach the singularity of \ref{eq:beta_TM} and we can
approximately write the normal component of the wave vector in the
uniaxial material slab for obliquely incident plane waves as \e
\beta_{\rm TM} \approx jk_{\rm t}\sqrt{\frac{\varepsilon_{\rm
t}}{|\varepsilon_{\rm n}|}}. \label{eq:beta_TM_appr}\f
For the
normal incidence ($k_{\rm t}=0$) the incident plane wave has no
electric field component along the normal and
therefore the normal component of the permittivity has no effect.
We can now write the surface impedance of
the grounded uniaxial slab for oblique incidence using
\ref{eq:Z_s^TM} and \ref{eq:beta_TM_appr} as: \e Z_{\rm s}^{\rm
TM} = -j\frac{k_{\rm t}}{\omega\sqrt{\varepsilon_{\rm
t}|\varepsilon_{\rm n}|}}\tanh\left(k_{\rm
t}\sqrt{\frac{\varepsilon_{\rm t}}{|\varepsilon_{\rm n}|}}h \right).
\label{eq:Z_s^TM_1}\f
For TM polarization inductive surfaces
support surface waves, because the surface impedance for
the evanescent TM-polarized plane wave in free space is
$-j\frac{\eta_0\gamma_{\rm n}}{k_0}$, where $\gamma_{\rm n} =
\sqrt{k_{\rm t}^2 - k_0^2}$ and $\eta_0$ is the free-space wave
impedance. We see from \ref{eq:Z_s^TM_1} that in this case the
surface does not support TM-polarized surface waves, and its surface
impedance is not a periodic function of the frequency. More importantly, we see that
as $\varepsilon_{\rm n}$ approaches zero, the surface impedance
for the TM polarization $Z_{\rm s}^{\rm TM}$ tends to $(-j\infty)$.

For the case when $\varepsilon_{\rm n} \rightarrow (-0)$  we can
write the surface impedance \ref{eq:Z_s^TM} as: \e Z_{\rm s}^{\rm
TM} = j\frac{k_{\rm t}}{\omega\sqrt{\varepsilon_{\rm
t}|\varepsilon_{\rm n}|}}\tan\left(k_{\rm
t}\sqrt{\frac{\varepsilon_{\rm t}}{|\varepsilon_{\rm n}|}}h \right).
\label{eq:Z_s^TM_2}\f We see that when $\varepsilon_{\rm n}
\rightarrow (-0)$ the waves do not become evanescent in the uniaxial
material slab and the surface impedance
\ref{eq:Z_s^TM_2} becomes a quickly oscillating function of $\omega$. The surface
operates as a magnetic conductor either for a narrow frequency band
or for a limited range of incidence angles. However, we can overcome this
drawback by choosing the tangential component of the permittivity to
be negative instead of $\varepsilon_0$ and make the fields
propagating along the normal of the surface evanescent in the
material slab. For negative values of the permittivity and in cases
when $|\varepsilon_{\rm t}| \gg |\varepsilon_{\rm n}|$, we can
replace the tangent with the hyperbolic tangent and
$\varepsilon_{\rm t}$ with its absolute value in
\ref{eq:Z_s^TM_2}.

Let us next look at the case of electrically thin material slabs.
In this case we can replace the tangent function by its argument.
Choosing $|\varepsilon_{\rm t}| \rightarrow 0$ and
$\varepsilon_{\rm n}=\varepsilon_0$ we make the uniaxial material slab electrically thin.
In this case a near-zero value of the transversal permittivity is not needed
for realization of a high surface impedance: \e Z_{\rm s}^{\rm TM}
\approx j\omega\mu_{\rm t}h - j\frac{k_{\rm
t}^2}{\omega\varepsilon_{\rm n}}h\label{eq:Z_s^TM_3}.\f Instead
a high-$\mu_{\rm t}$ or a low-$\varepsilon_{\rm n}$ material is
needed. In terms of surface wave suppression, the latter choice
outperforms the former. Unfortunately, the latter choice will lead
to a situation where the performance of the magnetic conductor
deteriorates in the vicinity of the incidence angle defined by \e
\sin^2\left(\theta_1\right) = \mu_{\rm t}\varepsilon_{\rm n}c_0^2,
\f where $\theta_1$ is the angle of incidence and $c_0$ is the speed
of light in free space. In the case $\mu_{\rm t} \rightarrow \infty$ the approximation of
the electrically thin substrate results in the following conditions for the transverse
components of the permittivity: \e \frac{\varepsilon_{\rm
t}}{\varepsilon_0} \ll \frac{\mu_0}{\mu_{\rm t}}, \hspace{1cm}
\varepsilon_{\rm t} \ll \varepsilon_{\rm n}.\f
These conditions are difficult to satisfy because of the complexity of the needed
uniaxial grounded material slab structure. First, we would need to
have metallic wires (for near-zero values of the relative
permittivity) in all three directions. Secondly, we would need to add
split-ring resonators or similar inclusions (for high values of relative permeability). We can
conclude that the regime of electrically thin substrate leads to the
situation where either the performance of the magnetic conductor
deteriorates at a certain incidence angle or the realization of the
magnetic conductor is difficult.

In the case of a very high $\mu_{\rm t}$ (the normal component of $\mu$ has no effect on
the incident wave in the TM case), most of the incident wave power is refracted along
the normal. Then the response of the
slab becomes weakly dependent on the incidence angle, since \e Z_{\rm
s}^{\rm TM} \approx j\sqrt{\frac{\mu_{\rm t}}{\varepsilon_{\rm
t}}}\tan\left(\omega\sqrt{\varepsilon_{\rm t}\mu_{\rm t}}h\right).
\label{eq:Z_s^TM_4}\f In addition to this, the slab becomes
electrically thick which results in quickly oscillating (with respect
to the frequency) surface impedance.

From the discussion above we see that it is possible to realize artificial magnetic
conductors for TM polarization by having one of the components of $\varepsilon$ close to zero.
$\varepsilon_{\rm n} \rightarrow (+0)$ is a suitable choice
if we tolerate the loss of the magnetic wall effect at the angle $\theta_1=0$. Realization of
the high surface impedance over a wide frequency band or over a wide range of
incidence angles otherwise is not possible without specially adjusting the value of one of the three remaining material parameters.

\section{Magnetic conductors for TE polarization}\label{Magnetic conductors for TE polarization}

For the TE polarization the case of the grounded uniaxial material slab becomes different than for its counterpart TM polarization discussed in the previous section. The surface impedance in \ref{eq:Z_s^TE} is more dependent on the permeability rather than permittivity. In order to realize a magnetic conductor from an isotropic material, the value of the permittivity needs to be minimized and the value of the permeability maximized (see also \ref{eq:Z_s^TE} and \ref{eq:Z_s^TM}). For this reason the singularity of \ref{eq:beta_TE} does not play so significant role as it does for the TM polarization. Instead, we concentrate on the regime $\mu \gg 1$.

In order to minimize the angular dependency of the magnetic conductors, we should have
$\mu_{\rm n} \gg 1$. Then we have from \ref{eq:beta_TE} and \ref{eq:Z^TE}: \e Z_{\rm s}^{\rm TE} \approx j\sqrt{\frac{\mu_{\rm
t}}{\varepsilon_{\rm t}}}\tan\left(\omega\sqrt{\varepsilon_{\rm
t}\mu_{\rm t}}h\right). \f For a high surface impedance we still
need to maximize $\mu_{\rm t}$, which unfortunately leads to quick oscillations of $Z_{\rm s}^{\rm TE}$ along the frequency axis. We can overcome this drawback by decreasing the transverse
permittivity and thus complicating the layer structure.

We can also decrease the angular dependency of the surface impedance
by having the transverse permittivity approach zero:
$\varepsilon_{\rm t} \rightarrow 0$. In this case the surface
impedance of TE polarized incident waves reads \e Z_{\rm s}^{\rm TE}
\approx j \frac{\omega\sqrt{\mu_{\rm t}\mu_{\rm n}}}{k_{\rm
t}}\tanh\left(k_{\rm t}\sqrt{\frac{\mu_{\rm t}}{\mu_{\rm n}}}h
\right).\f Obviously we still need to maximize the transverse
permeability in order to realize a magnetic conductor.

The magnetic wall effect realized by using grounded uniaxial
material slabs for the TE polarization does not seem to offer such
advantages as for the TM polarization. For the TM polarization it is
sufficient to engineer only the normal component of the effective
permittivity correctly in order to realize a magnetic conductor. The
case of the TE polarization is not so simple; for the TE
polarization at least two effective material parameters need to be
engineered in the transverse direction. Practically this means
orienting two different type of particles (magnetic and electric) in two orthogonal directions.


\section{Practically realizable uniaxial magnetic conductors}\label{Practically realizable uniaxial magnetic conductors}

\begin{figure}[t!]
\centering
\subfigure[]{\includegraphics[width=4cm]{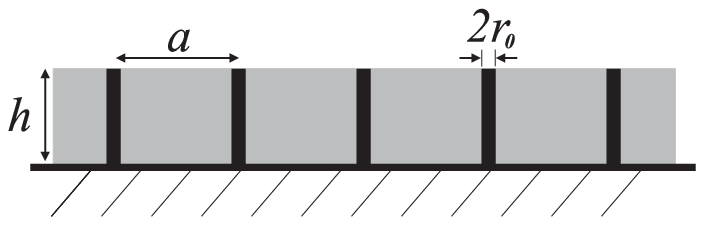}}
\subfigure[]{\includegraphics[width=3.5cm]{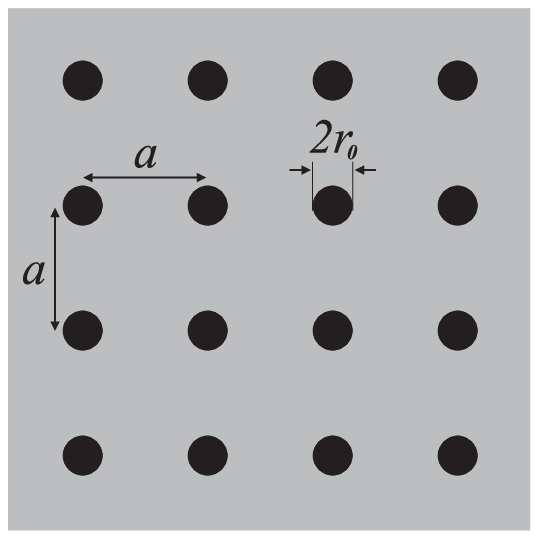}}
\caption{ (a) A side  and (b) a top view of the spatially dispersive
wire medium slab. The metallic vias having a diameter $2r_0$ are
separated from each other by a period $a$. The vias are embedded in
a dielectric substrate.  The height of the structure is $h$ and the
substrate permittivity is $\varepsilon_{\rm h}$. } \label{fig:3}
\end{figure}

The wire medium serves as a good example of a uniaxial
electromagnetic crystal that can possess near zero values of the
effective permittivity. Based on the discussion in the previous
sections the most promising case is to use this feature for
realizing magnetic conductors for TM polarization. In this case the wires should be oriented normally
to the surface. For TE-polarized incident fields these wires
have no effect. This type of grounded wire medium slab structures,
or "Fakir's bed of nails", have been studied earlier in
\cite{king,mario} (see Fig.~\ref{fig:3}). In \cite{king} the wire
medium slab comprised densely packed wires, and the
height of the slab was large compared to the period of
the wire medium. The surface impedance for such a structure
reads: \e Z_{\rm s}^{\rm TM} = j\sqrt{\frac{\mu_{\rm
t}}{\varepsilon_{\rm t}}}\tan\left(\omega\sqrt{\varepsilon_{\rm
t}\mu_{\rm t}}h \right),\f as was shown also in \cite{mario}. For
this type of structures we see that the near-zero value of the
normal component of the permittivity tensor has no effect on the
surface impedance, but the TEM modes of
the wire medium become dominant (see, e.g. \cite{Tretyakov}, p.~231). Furthermore, structures with densely packed metallic vias are expensive to manufacture. Instead, thin slabs with rather sparse wires are preferred in practical applications. In this case also the TM modes excited in the wire medium slab become significant. In the following, we will discuss the structure
needed for realizing the wide-angle and wide-band magnetic conductors for the TM polarization. However, in practical applications the bandwidth of the magnetic conductor is reduced by the limited frequency band for which the wire medium posses near-zero values for the effective permittivity.

The wire medium is known to be strongly spatially dispersive material
whose effective permittivity can be described for our case as
\cite{Shvets,Belov,Simovski2}: \e \varepsilon_{\rm n} =
\varepsilon_0\varepsilon_{\rm h}\left(1 - \frac{k_{\rm p}^2}{k^2 -
\beta^2} \right), \label{eq:eps_n_spatial}\f where
$\varepsilon_{\rm h}$ is the relative permittivity of the host
medium, $k_{\rm p}$ is the plasma wave number,
$k=k_0\sqrt{\varepsilon_{\rm h}}$ is the wave number in the host
medium, and $\beta$ is the $z$-component of the wave vector.
This non-local homogenization model predicts that in the case of
TM-polarized incident waves two modes can be excited in the wire
medium with different values for $\beta_{\rm z}$. In this
case the definition of the $\varepsilon_{\rm n}$ for the wire medium
slab becomes non-explicit and the reflection properties of such
slabs need to be solved by calculating the fields of both modes in
the slab \cite{mario}. In addition to this, the study in the
previous sections becomes not applicable in the spatially dispersive case. For these reasons we need to suppress the spatial dispersion in the wire medium slab.

\begin{figure}[t!]
\centering
\subfigure[]{\includegraphics[width=4cm]{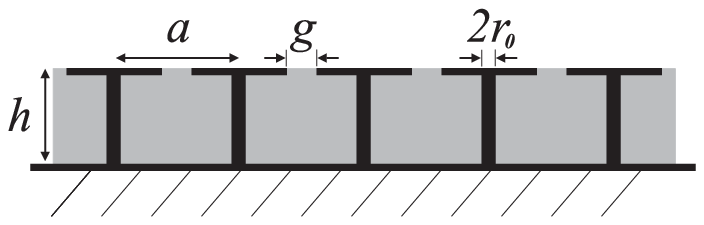}}
\subfigure[]{\includegraphics[width=3.5cm]{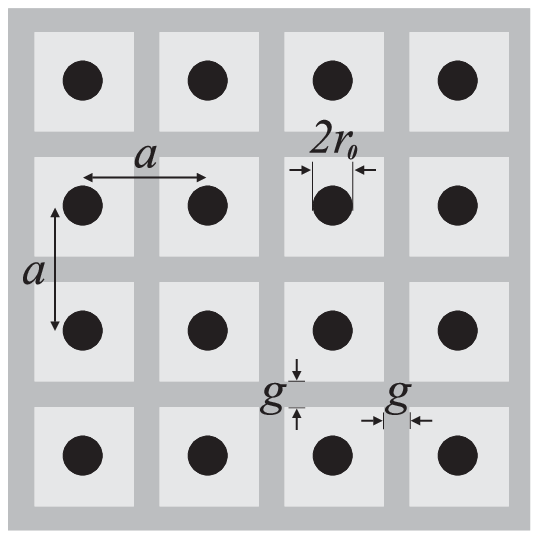}}
\caption{ (a) A side  and (b) a top view of the non-spatially
dispersive wire medium slab. The metallic vias having a diameter
$2r_0$ are separated from each other by a period $a$ and each via is
galvanically connected to a metallic patch. The spacing between the
adjacent patches is $g$. The height of the structure is $h$ and the
substrate permittivity is $\varepsilon_{\rm h}$. } \label{fig:4}
\end{figure}

Suppression of the spatial dispersion in wire medium has been
studied recently in \cite{Demetriadou, Olli, Yakovlev, Yakovlev2}.
In \cite{Demetriadou} the spatial dispersion was suppressed by
connecting the parallel wires capacitively to each other. This is
similar to the situation in \cite{Olli,Yakovlev,Yakovlev2} where it
was first noticed that the spatial dispersion in the wire medium
slab covered with metallic patches is suppressed when choosing the
height and the lattice constant of the wire medium slab properly. In
these cases the local (non-spatially dispersive) model is valid for
the effective value of the normal component of the permittivity, and
it reads: \e \varepsilon_{\rm n} = \varepsilon_0\varepsilon_{\rm
h}\left(1 - \frac{k_{\rm p}^2}{k^2} \right).
\label{eq:eps_n_nonspatial}\f The plasma wave number $k_{\rm p}$ can
be estimated as \cite{Belov}: \e k_{\rm p}^2 =
\frac{2\pi}{a^2\left(\ln\left( \frac{a}{2\pi r_0}\right) +
0.5275\right)} \label{eq:kp2},\f where $a$ is the lattice constant
of the wire medium and $r_0$ is the radius of the vias (see
Fig.~\ref{fig:4}).

The explanation of the suppression of spatial dispersion in the wire
medium is simple. The presence of patches makes the currents induced
in the wires by the incident TM-polarized waves practically uniform
over the height $h$. It is so if $h < a$, since in this case the
induced charges are practically concentrated at the patches' edges.
The absence of charge accumulation on the wires ensures the
practical uniformity of currents. Though mathematically the currents
along the wires can still be presented as a sum of the TEM and TM
modes, the smallness of the current variation across the wire medium
slab means that the propagation effect of these modes over the
height $h$ is negligible. Since the waves in the wire medium slab
can propagate only in the horizontal plane, one can consider the
wire medium slab as a uniaxial dielectric with negative
$\varepsilon_{\rm n}$ below the plasma frequency.

\section{An example and numerical validation}\label{An example and numerical validation}
\begin{figure}[t!]
\centering \epsfig{file = 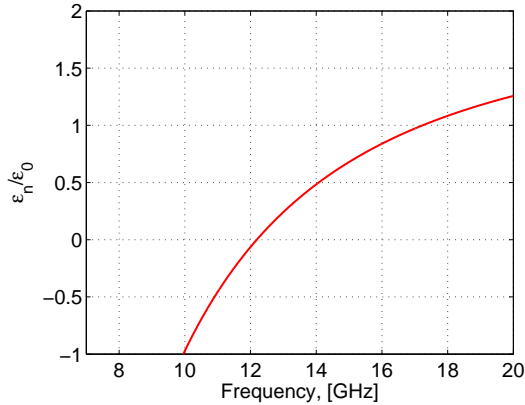, width = 7.5cm} \caption{The
normal component of the relative permittivity of the wire medium
slab calculated using \ref{eq:eps_n_nonspatial}.} \label{fig:5}
\end{figure}
\begin{figure}[t!]
\centering \subfigure[]{\epsfig{file = 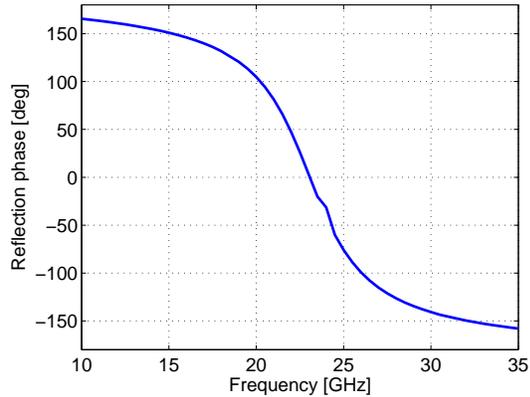, width =
7.5cm}} \subfigure{\epsfig{file = 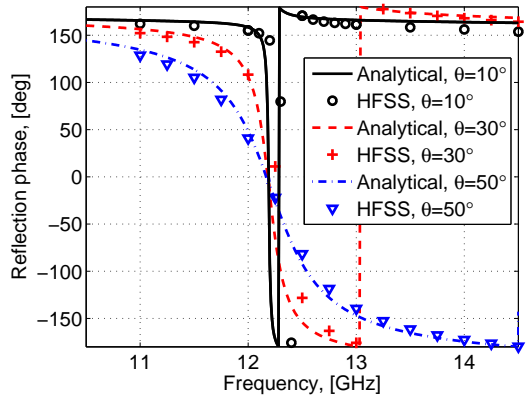, width = 7.5cm}}
\caption{(a) The simulated reflection phase for the normal
incidence.
(b) The reflection phase for oblique incidences of 10, 30, and 50
degrees. The analytical results are calculated using
\ref{eq:Z^TM}. For the normal incidence the vias do not have any
effect on the plane wave response of the structure.} \label{fig:6}
\end{figure}
Here we concentrate on validating our results for the TM
polarization using an example that follows the design of
\cite{Olli,Yakovlev,Yakovlev2}. In our example shown in
Fig.~\ref{fig:4} we have included metallic patches on top of the
wire medium slab so that each patch is galvanically connected to
just one via. The structure has similar appearance to the mushroom
structure. However, here we do not utilize the structure resonance.
Instead, we utilize the near-zero value of the normal component of
the permittivity of the wire medium slab in order to create a
magnetic conductor. To demonstrate that the magnetic conductor
effect of the surface relates to the near-zero value of the normal
component of the permittivity we have used qualitatively the design
equations of \cite{Luukkonen1}. These equations allow us to design
the resonance frequency of the structure (the mushroom type of
high-impedance surfaces) far away from the frequency region where we
would have $\varepsilon_{\rm n} \rightarrow 0$.

We have chosen the parameters of the non-spatially dispersive wire
medium slab to be the following: $a=4$\,mm, $g=0.5$\,mm,
$h=0.5$\,mm, $\varepsilon_{\rm h} = 2$, and $r_0 = 0.05$\,mm. For
these parameters the normal component of the effective permittivity
tends to zero approximately at $12.2$\,GHz and the mushroom
structure resonance is approximately at $23$\,GHz as shown in
Figs.~\ref{fig:5} and \ref{fig:6}(a), respectively. For the normal
incidence the vias do not have any effect on the plane-wave response
of the structure, and the resonance shown in Fig.~\ref{fig:6}(a) is
solely due to the other structural properties of the wire medium
slab, that is, the effective capacitance between the adjacent
patches and the inductive response of the grounded dielectric slab.
The results for the reflection phase have been verified using
Ansoft's High Frequency Structure Simulator (HFSS) \cite{HFSS}. We
can therefore explicitly distinguish the magnetic conductor effect
due to $\varepsilon_{\rm n} \rightarrow 0$ from the known parallel circuit response of a high-impedance surface \cite{Sievenpiper,Monorchio,Luukkonen1}.

\begin{figure}[t!]
\centering \subfigure[]{\includegraphics[width=3.7cm]{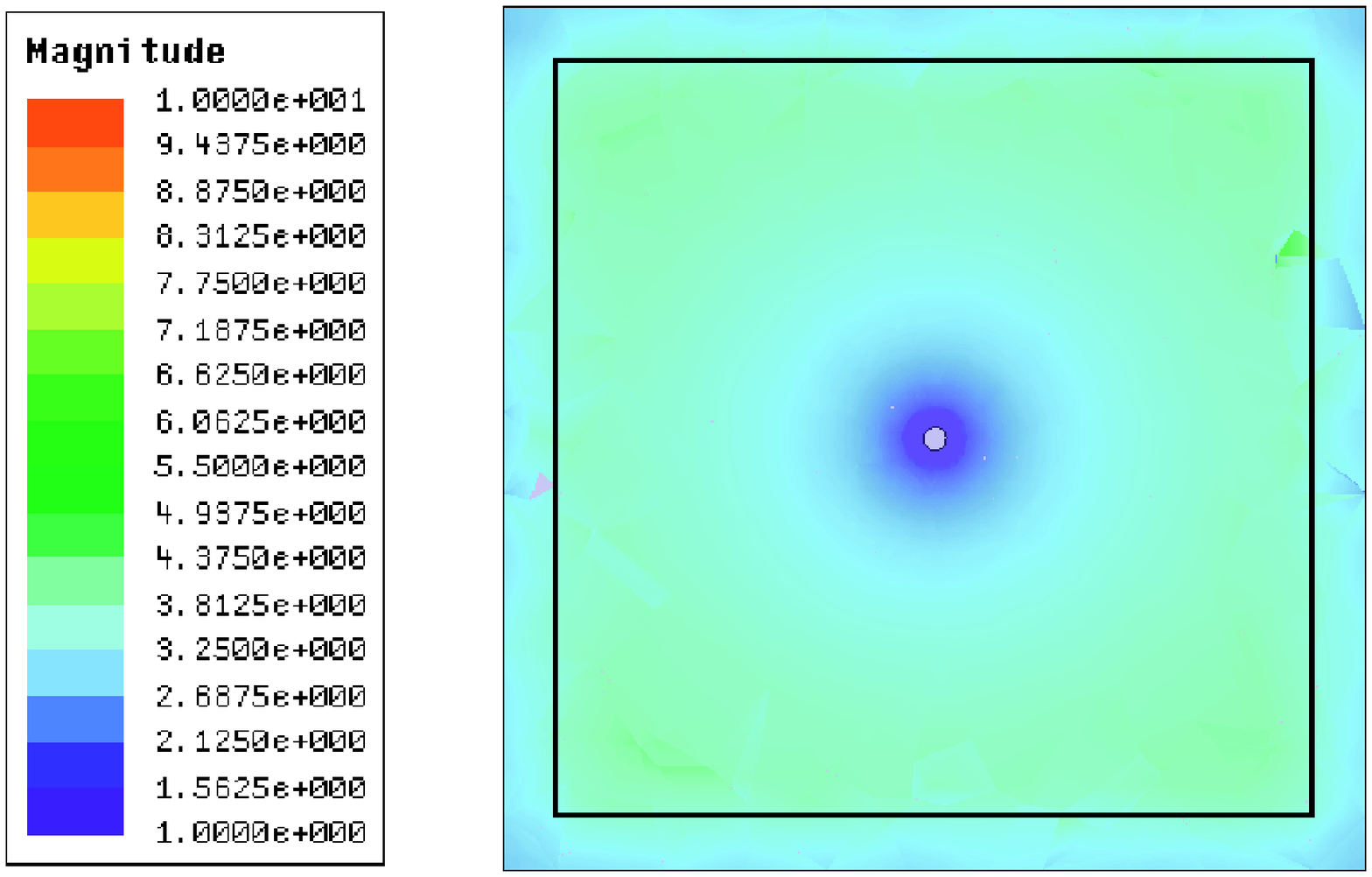} }
\subfigure[]{\includegraphics[width=2.33cm]{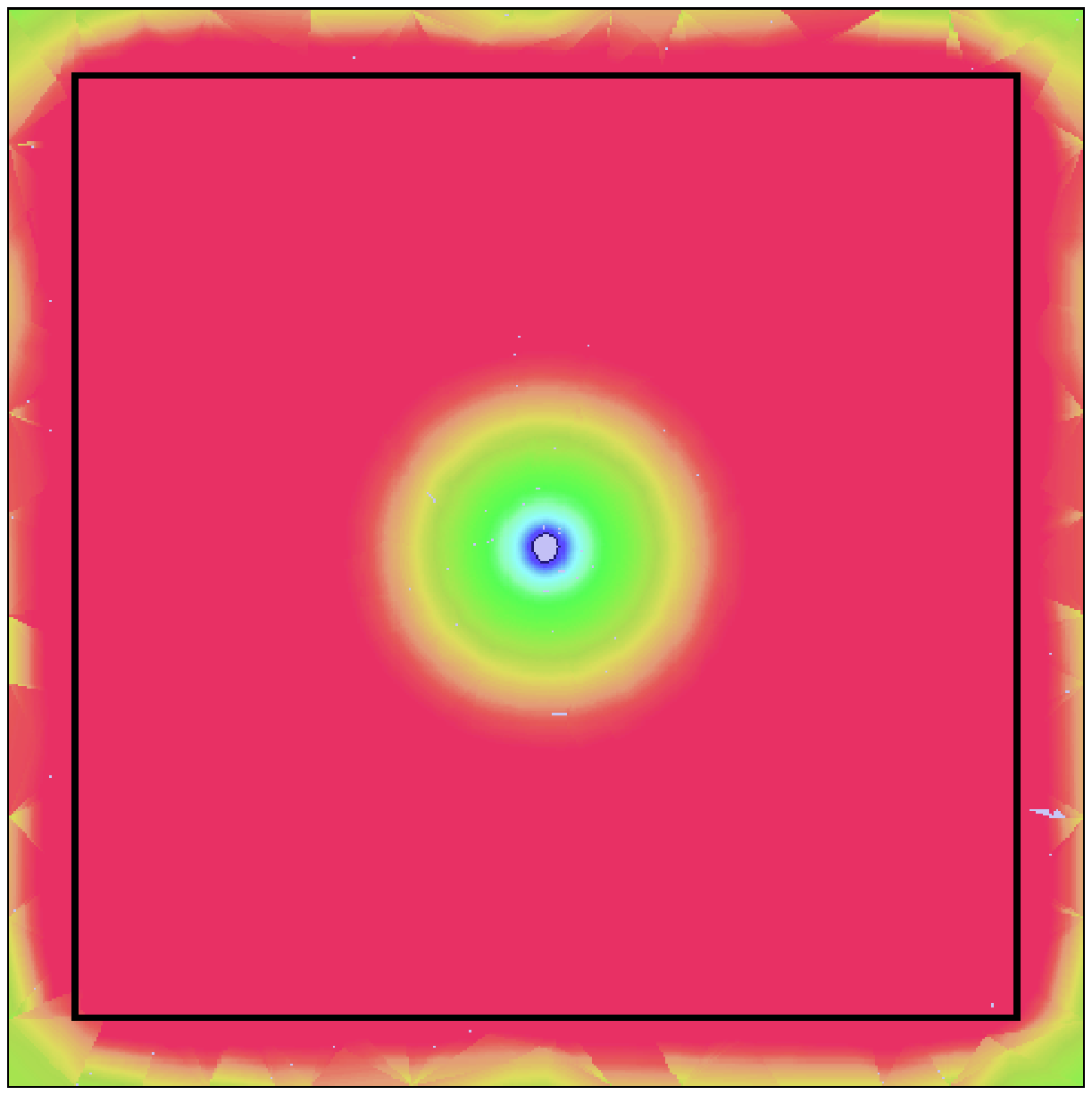}}
\subfigure[]{\includegraphics[width=2.33cm]{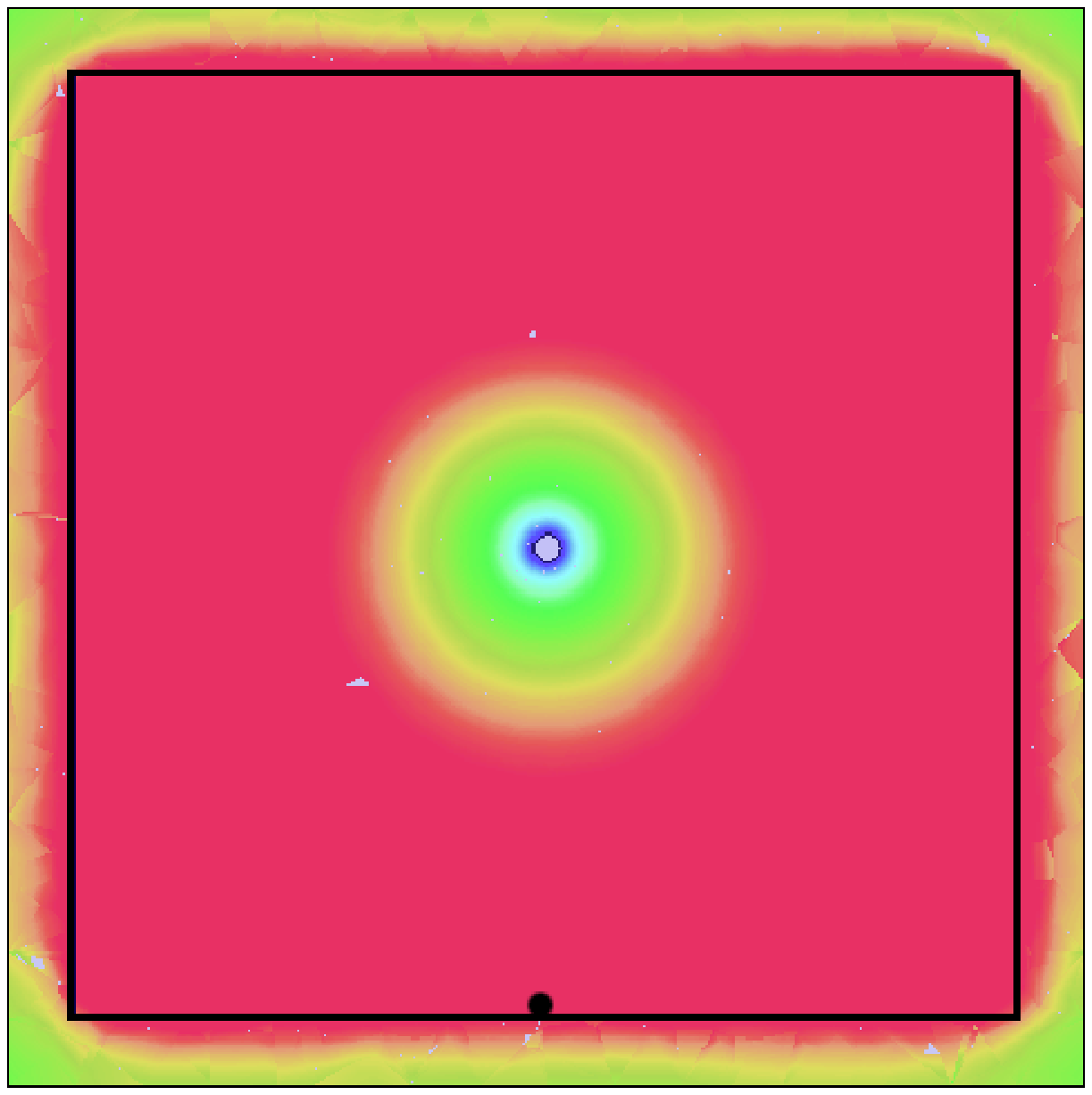}} \caption{The
magnitude of the normal component of the electric field magnitude,
$E_{\rm n}$, over one unit cell at 12\,GHz for the incidence angles
of (a) 10, (b) 30, and (c) 50 degrees. The magnitudes have been
plotted on a surface 0.25\,mm above the ground plane.} \label{fig:7}
\end{figure}

\begin{figure}[t!]
\centering \subfigure[]{\includegraphics[width=3.7cm]{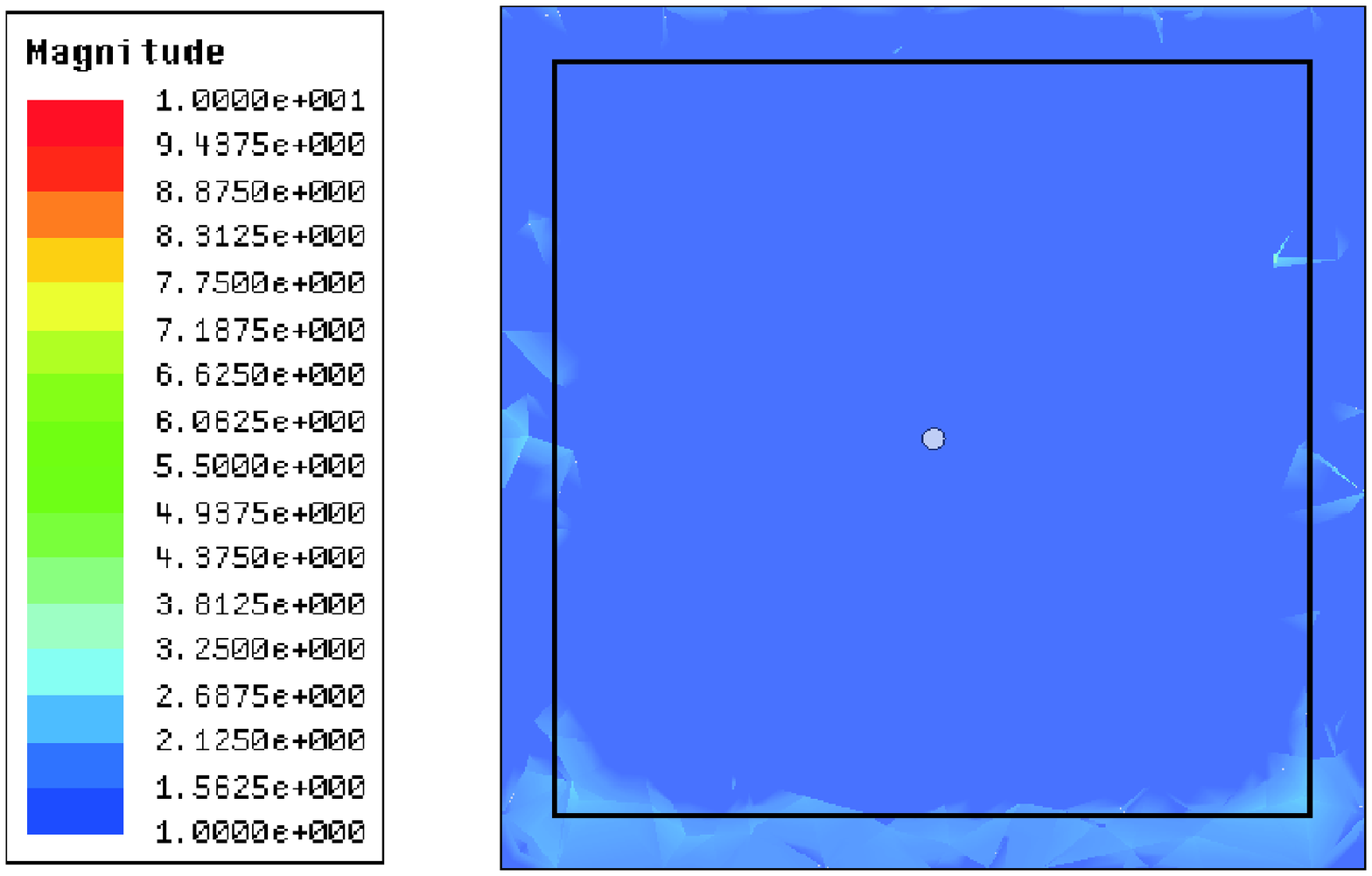} }
\subfigure[]{\includegraphics[width=2.35cm]{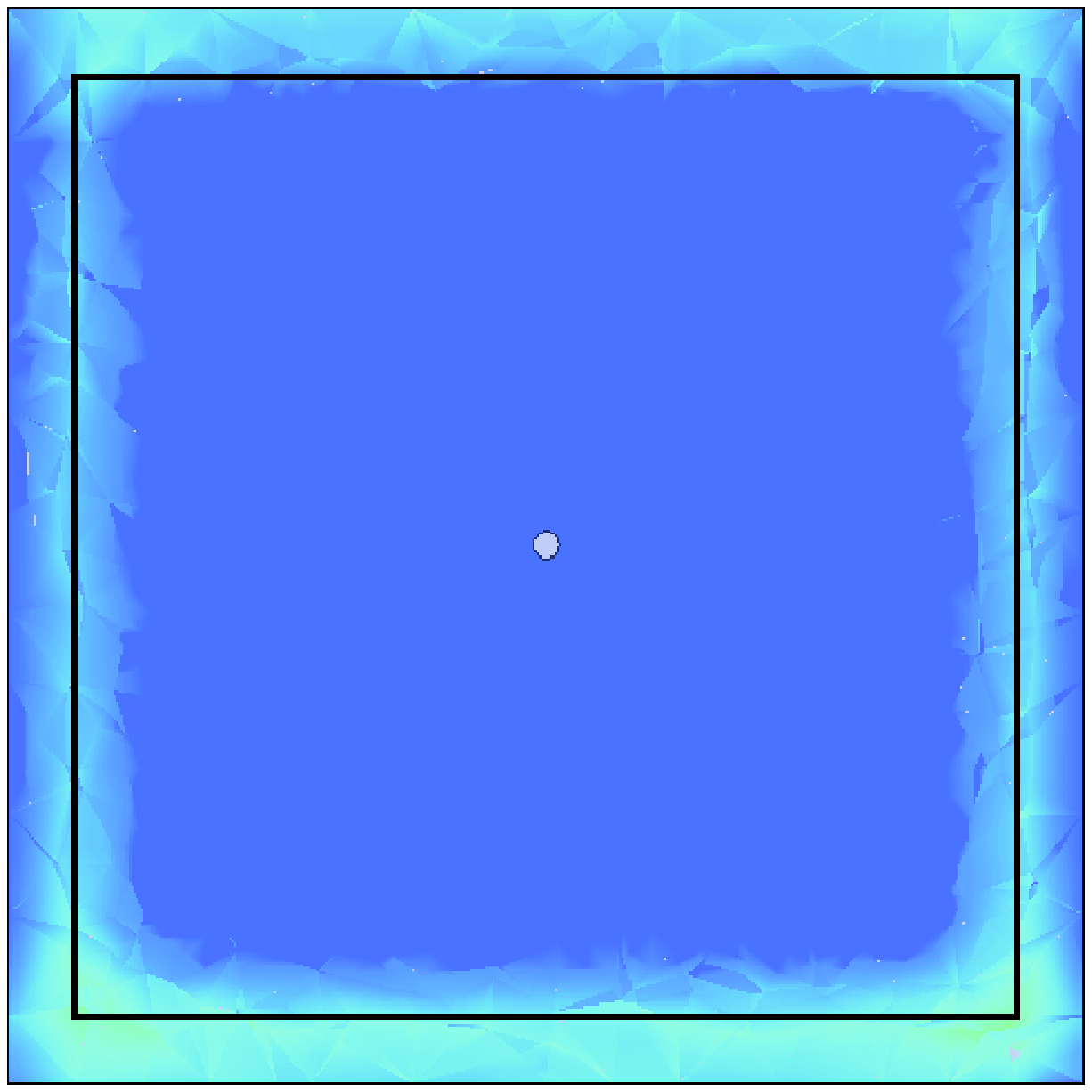}}
\subfigure[]{\includegraphics[width=2.35cm]{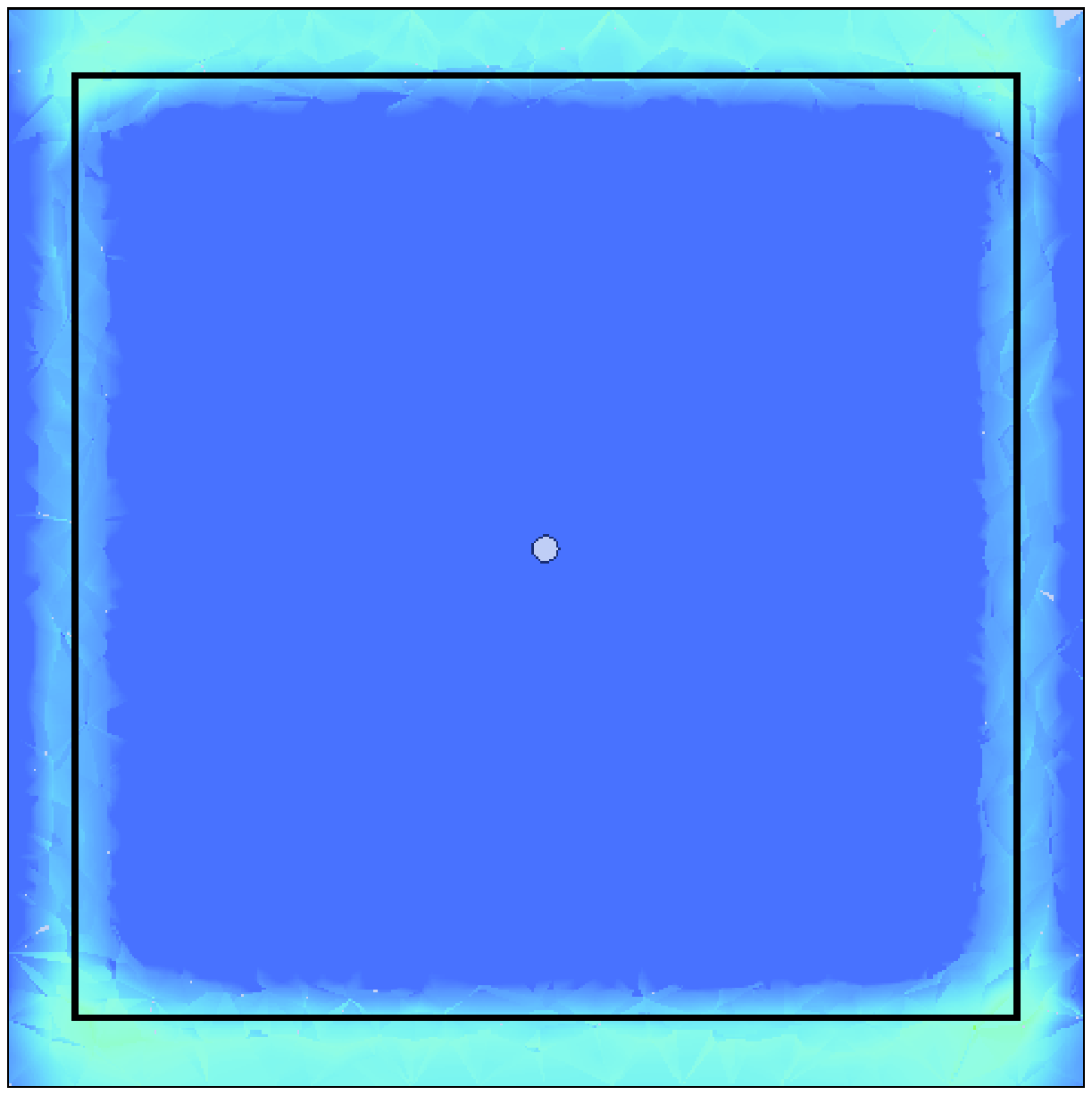}} \caption{The
magnitude of the tangential component of the electric field
magnitude, $E_{\rm t}$, over one unit cell at 12\,GHz for the
incidence angles of (a) 10, (b) 30, and (c) 50 degrees. The
magnitudes have been plotted on a surface 0.25\,mm above the ground
plane.} \label{fig:8}
\end{figure}

For oblique incidence the wire medium is excited by incident  plane
waves. The main result which we obtain using this description of
the metasurface is the dramatic reduction of the resonance frequency
for oblique incidence starting from very small angles. In
Fig.~\ref{fig:6}(b) the reflection phase for the incidence angles of
10, 30, and 50 degrees calculated using \ref{eq:Z^TM} is shown. We
see clearly that the surface behaves as a magnetic conductor
(reflection phase equals to zero) at approximately $12.2$\,GHz for
all of the incidence angles because $\varepsilon_{\rm n} \rightarrow
0$ as predicted by \ref{eq:Z_s^TM_1}. The HFSS simulations agree
well with our analytical results. However, for the incidence angle
of 10 degrees the difference between the analytical and the
simulated results is somewhat larger than for more oblique
angles. Furthermore, Eq.~\ref{eq:eps_n_nonspatial} implies
that the electric fields inside the wire medium slab are mainly
oriented along the metallic vias. Indeed, this can be confirmed by
looking at the simulated magnitudes of the normal and transverse
components of the electric field shown in Figs.~\ref{fig:7} and
\ref{fig:8}, respectively, for different angles of incidence at
12\,GHz. The magnitudes of the electric field components have been
plotted at $0.25$\,mm above the ground plane. We see also that for
the incidence angle of 10 degrees the electric field lines are not
as parallel to the metallic vias as they are for the other cases.
Because of this, the frequency for which $\varepsilon_{\rm n} = 0$
is not exactly that predicted by \ref{eq:eps_n_nonspatial}. This
explains the larger difference between the analytical and simulated
results for the incidence angle of 10 degrees compared to the larger
incidence angles in Fig.~\ref{fig:6}(b).

\section{Discussion and conclusions}\label{Discussion and conclusions}

In this paper we have studied the possibility to use grounded
uniaxial materials as magnetic conductors for both TE and TM fields.
We have first studied the properties of such surfaces through the
derived surface impedance expressions. Attention has been paid
especially to the TM polarization which is more promising in view of practical realizations. Based on the results of the theoretical studies we have designed an artificial magnetic conductor having the appearance of a Sievenpiper mushroom structure based on wire-medium slab,
but operating in a different regime.
The epsilon-near-zero resonance which we use here to
create the magnetic conductor effect takes place at a frequency
which is approximately twice as lower as the resonant frequency of the same
structure at the conventional parallel resonance. This offers a
possibility for significant reduction of the layer thickness or for a
dual-frequency design, as proposed in \cite{absorber}.
The magnetic wall effect is very little affected by the
incidence angle (excluding $\theta_1\approx0$). We have verified our
results with simulations.

The present study of the properties of grounded uniaxial materials
relates closely to the recent studies on the electromagnetic
boundary conditions for anisotropic metamaterials
\cite{Lindell,Lindell1}, where instead of treating the transversal
components of the incident fields, the authors propose boundary
conditions for the normal components of the incident fields. In
\cite{Lindell,Lindell1} the authors arrive to the same conclusion as
the results of this paper, that the perfect magnetic boundary
condition can be achieved for the TM field by having
$\mathbf{n}\cdot \mathbf{D}=0$. In the present paper this is proposed to realize
with wire medium at its plasma frequency. It is clear that this
boundary condition differs considerable from the conventional
boundary condition for the perfect magnetic conductors
$\mathbf{n}\times \mathbf{H} =0$.

The magnetic wall effect realized for the TM-polarization using the wire
medium is limited to the vicinity of the plasma frequency of the
wire medium. Because of this, the bandwidth of the magnetic wall
effect is small, especially for small incidence angles. However, in
some applications, such as GPS or GALILEO, this does not
impose any limitations as the needed frequency band for these
applications is rather small. In other applications the proposed novel
way of realizing the surface resonance can be used to create a
secondary resonance in the vicinity of the primary resonance
\cite{absorber}. Also, the surface does not operate as an AMC for
the normal incidence. In a practical environment this should not be
a problem as the waves impinging to the surface are seldom just
normally oriented. Further, in AMC antenna applications, where the
primary radiator lies very close to the AMC surface, most of the
power from the primary radiator is radiated in the angular spectrum
practically excluding the normal incidence.


\section*{Acknowledgements}\label{Acknowledgements}

The discussions with Prof. Igor Nefedov are warmly acknowledged.
Olli Luukkonen wishes to thank the Nokia Foundation, Jenny and Antti
Wihuri Foundation, Emil Aaltonen Foundation, and the Finnish Society
of Electronics Engineers for financial support.


\end{document}